\documentclass{article}
\usepackage{arxiv} 
\usepackage[utf8]{inputenc} 
\usepackage[T1]{fontenc}    
\usepackage{hyperref}       
\usepackage{url}            
\usepackage{booktabs}       
\usepackage{amsfonts}       
\usepackage{nicefrac}       
\usepackage{microtype}      
\usepackage{graphicx}       
\usepackage{amsmath}        

\title{Improving SSVEP BCI Spellers with Data Augmentation and Language Models}

\author{%
  Joseph Zhang \\
  Biomedical Engineering\\
  Carnegie Mellon University\\
  Pittsburgh, PA 15213 \\
  \texttt{josephzh@andrew.cmu.edu} \\
  \And
  Ruiming Zhang \\
  Biomedical Engineering\\
  Carnegie Mellon University\\
  Pittsburgh, PA 15213 \\
  \texttt{ruiming2@andrew.cmu.edu} \\
  \And
  Kipngeno Koech \\
  College of Engineering \\
  Carnegie Mellon University Africa\\
  BP 6150, Kigali, Rwanda\\
  \texttt{bkoech@andrew.cmu.edu} \\
  \And
  David Hill \\
  Robotics Institute \\
  Carnegie Mellon University \\
  Pittsburgh, PA 15213 \\
  \texttt{davidhil@andrew.cmu.edu} \\
  \And
  Kateryna Shapovalenko \\
  Language Technologies Institute \\
  Carnegie Mellon University \\
  Pittsburgh, PA 15213 \\
  \texttt{kshapova@andrew.cmu.edu} \\
}

\date{December 27, 2024}

\begin{document}

\maketitle

\begin{abstract}
  Steady-State Visual Evoked Potential (SSVEP) spellers are a promising communication tool for individuals with disabilities. This Brain-Computer Interface utilizes scalp potential data from (electroencephalography) EEG electrodes on a subject's head to decode specific letters or arbitrary targets the subject is looking at on a screen. However, deep neural networks for SSVEP spellers often suffer from low accuracy and poor generalizability to unseen subjects, largely due to the high variability in EEG data. In this study, we propose a hybrid approach combining data augmentation and language modeling to enhance the performance of SSVEP spellers. Using the Benchmark dataset from Tsinghua University, we explore various data augmentation techniques, including frequency masking, time masking, and noise injection, to improve the robustness of deep learning models. Additionally, we integrate a language model (CharRNN) with EEGNet to incorporate linguistic context, significantly enhancing word-level decoding accuracy. Our results demonstrate accuracy improvements of up to 2.9 percent over the baseline, with time masking and language modeling showing the most promise. This work paves the way for more accurate and generalizable SSVEP speller systems, offering improved communication solutions for individuals with disabilities.
\end{abstract}

\noindent \textbf{Keywords:} SSVEP, Brain-Computer Interface, EEG, Deep Learning, Letter Frequency, Language Models, RNN 

\section{Introduction}

Brain-Computer Interfaces (BCIs) are systems that allow direct communication between the brain and an external device. BCIs have transformative potential, particularly for individuals with severe physical disabilities, offering applications in communication, environmental control, and mobility. Our group is primarily interested in steady-state visually-evoked potential (SSVEP), which is a neural signal that occurs when a person looks at flickering stimuli at specific frequencies. This signal is commonly used in a speller task, where a user will try to spell words by looking at a computer screen with a matrix of characters each flickering at specific frequencies, and their SSVEP data is recorded using electroencephalography (EEG). This data is preprocessed and then decoded to determine what the user was looking at, and recently, the use of deep neural networks (DNNs) for decoding SSVEP has become more viable and effective, specifically the use of EEGNet, a CNN that is commonly used in EEG research. However, this process is still prone to errors due to the noisy and highly variable nature of EEG data. One especially difficult area is generalization, as EEG-based DNNs tend to not generalize well into new data. Therefore, our group experimented with various components of the BCI data pipeline to improve the efficacy of SSVEP DNN decoders.

To improve SSVEP BCI performance, we explored data augmentation and language modeling. Data augmentation aims to increase the accuracy and stability of a model by creating more data to be used in training which is derived from the original dataset. This new data can be transformed from the dataset using a variety of functions. For our project, we plan to combine the common augmentation of noise addition and Fourier domain random phase shifting to improve our baseline model's performance.

We also incorporated letter frequency into character classification as the real-world application of these speller tasks would involve writing words and phrases which can be predicted using prior inputs. A character-based Recurrent Neural Networks (RNN) would dynamically adjust EEGNet’s output, emphasizing linguistically-probable characters. The final result would be a hybrid model containing both EEGNet and our language model named "CharRNN", which would work in parallel to decode each letter observed in the SSVEP speller.

\section{Literature Review}

\begin{itemize}

\item \cite{lashgari2020} is a very useful overview paper of various augmentation techniques used in the field of deep learning for EEG. It reviews the common augmentations as well as their effectiveness on different types of problems for BCI-related tasks. It splits the types of data augmentation into six main categories: noise addition, GAN for generating new data, sliding or overlapping windowing, sampling, Fourier transform transformations, and recombination after segmentation. The authors discuss the current state of the art in each of these techniques and which techniques generally perform better for different tasks such as motor imagery, motor tasks, and visual tasks. We plan to use this paper as well as the papers referenced in the article to determine a new augmentation technique.

\item \cite{ding2024} introduces an innovative data augmentation technique known as EEG mask encoding (EEG-ME) for deep learning (DL)-based classification in SSVEP-based BCIs. This method addresses a common challenge in DL models for BCIs: overfitting caused by the limited availability of EEG data. The proposed EEG-ME method improves model generalization by masking portions of the EEG data, thereby encouraging the model to learn more robust features. To validate the effectiveness of EEG-ME, three distinct network architectures were employed: CNN-Former (a hybrid model integrating convolutional neural networks and Transformer), time-domain-based CNN (tCNN), and a lightweight architecture known as EEGNet. Testing was conducted on publicly available benchmark and BETA datasets. The study demonstrates that EEG-ME substantially improves the average classification accuracy across these architectures and datasets.

\item \cite{graves2013} introduced a comprehensive framework for RNNs in sequence-based tasks, particularly emphasizing their ability to capture temporal dependencies over long sequences. By using a hidden state that evolves as the network processes each time step in the input, RNNs can retain information about previous elements, allowing for context-based predictions of characters, words, and phrases. For our project, we implemented a character-driven RNN to augment the classification probabilities of specific characters to aid the accuracy of EEGNet.

\item \cite{wan2023} presents GDNet-EEG, an attention-aware deep neural network based on group depth-wise convolution for recognizing SSVEP stimulation frequencies, with potential applications in early glaucoma diagnosis. The model employs group depth-wise convolution to extract temporal and spectral features from EEG signals across brain regions. It also integrates an attention mechanism to highlight essential brain regions for improved feature extraction. The model is validated on two public SSVEP datasets (Benchmark and BETA), as well as their combined dataset. With 1-second input signals, GDNet-EEG achieves average classification accuracies of 84.11\%, 85.93\%, and 93.35\% on the Benchmark, BETA, and combined datasets, respectively. Compared to baseline models, accuracies improve by 1.96\% to 18.2\%. This paper is especially important as we will use it to compare our baseline implementation.

\end{itemize}

The remaining literature was useful in researching approaches to improving SSVEP-based DNNs but ultimately was not used for our final implementations.

\begin{itemize}
\item \cite{guney2022} introduces a novel DNN architecture for steady-state visual evoked potential (SSVEP)-based brain-computer interfaces (BCIs) that significantly improves information transfer rates (ITRs) for BCI speller systems. The BCI speller task involves looking at various characters/symbols on a grid on a computer screen and attempting to spell words/phrases using just the input of sight, which is very beneficial to people with paralysis. The network involves multiple convolutional layers that captures the spatial and spectral information from the EEG signal. When tested on publicly-available Benchmark and BETA datasets, it reached ITRs of 265.23 bits/min on the Benchmark dataset and 196.59 bits/min on the BETA dataset, both achieved with just 0.4 seconds of stimulation.

\item \cite{liu2020} introduces the BETA dataset, a large-scale public Benchmark for SSVEP-BCIs, and it is used in works such as \cite{guney2022}. BETA includes EEG recordings from a 40-character speller using a range of flickering frequencies. The dataset contains multiple blocks of trials, with variations in signal quality due to real-world conditions, which makes it more representative of practical applications compared to controlled lab datasets. The paper also compares the performance of several frequency recognition methods (supervised and training-free) on the BETA dataset. Methods like TRCA, CCA, and FBCCA are evaluated to establish performance baselines for the SSVEP classification task. Although we do not use this dataset, it is commonly used in EEGNet model extensions that we researched.

\item \cite{zhang2024} makes progress in the field of BCI intention decoding for control of unmanned aerial vehicles (UAV's) and other intelligent vehicles using a lightweight network. There are three main paradigms or signals that have historically been used for this problem: motor imagery (MI), secure typing via BCI system with encrypted feedback (SSVEP), and average evoked potentials (ERP). This work uses both home-grown and public datasets with MI and SSVEP measurements for training, validating, and testing their model. Their network is composed of three main parts: a spatial feature extraction module (a 2D convolution followed by batch normalization), a temporal feature extraction module (a 1D convolution followed by batch normalization), and an attention feature enhancement module (noise adding, pooling, and fully connected layers). Finally, the output is flattened, fed through a few fully connected layers and a softmax layer for classification. 

\item \cite{nie2024} gives an overview of the challenges that multiclass classification poses in BCI deep learning. Good performance on datasets with more than 40 categories, such as the Benchmark dataset from Tsinghua University used in the paper, has not yet been achieved. To work towards this goal, the authors propose a novel analysis based on convolutional layers with residual connections. They achieve impressive results which they compare to other popular architectures trained on the same dataset. They achieve 52\% accuracy on a 0.2-second window and over 93\% accuracy on a 1-second window.

\item \cite{rostami2022} explores the enhancement of classification in real-world SSVEP data for BCI speller systems using deep convolutional neural networks (DCNNs). This study proposes using DCNNs to automatically extract relevant features from noisy SSVEP data to address the issue of noise and artifacts in the signal. The network was pre-trained using a portion of the BETA data and then retrained and tested at the individual subject level. Results indicated a significant improvement in both accuracy and information transfer rate (ITR) after retraining, with a p-value of less than 0.01 for all participants. Specifically, accuracy improved by 25.72 percent, and the ITR increased by 43.10 bits per minute (bpm).

\end{itemize}

\section{Baseline Model}

\begin{figure}[htbp]
    \centering
    \includegraphics[width=0.7\textwidth]{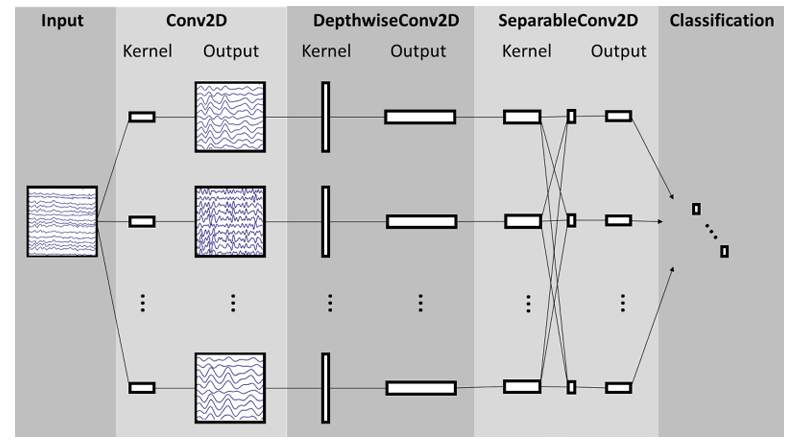} 
    \caption{Overall visualization of the EEGNet architecture\cite{lawhern2018}.}
    \label{fig:eegnet_architecture}
\end{figure}

EEGNet, developed by \cite{lawhern2018}, is a deep learning model specifically designed to analyze EEG data for BCI. As shown in Figure~\ref{fig:eegnet_architecture}, it is a compact convolutional neural network that uses depthwise and separable convolutions to extract EEG features and perform classification. EEGNet has shown strong performance in various EEG classification tasks, which makes it a reliable baseline for comparison with other models.

Fundamentally, the way EEGNet works is by leveraging convolutions to capture both the spatial and temporal aspects of the EEG data.  As shown in Table~\ref{table:eegnet_architecture}, EEGNet begins with a temporal convolution to capture frequency information. This is achieved by applying $F_1$ filters of size 1 x 64, where the filter length corresponds to half the EEG sampling rate, allowing it to capture frequency information above 2 Hz. The second block uses depth-wise convolution to learn spatial filters specific to each temporal filter. For each feature map from the previous layer, EEGNet applies a separate filter across the spatial dimension (channels), enabling EEGNet to learn spatial representations for each frequency band, like for instance, how alpha power is distributed along the cortex. This depth-wise convolution has $D$ x $F_1$ filters, where $D$ is the depth multiplier controlling the number of spatial filters per temporal filter. The next layer uses separable convolution, which combines depth-wise convolution (applied independently to each feature map) and point-wise convolution (applied across all feature maps) to reduce parameters and improve feature learning. Note that each layer includes batch normalization, dropout, and activation which is standard for CNNs. The final layer is a classification layer that assigns an array of probabilities for the various classes to the input signal.

\begin{table}[htbp]
    \centering
    \includegraphics[width=0.8\textwidth]{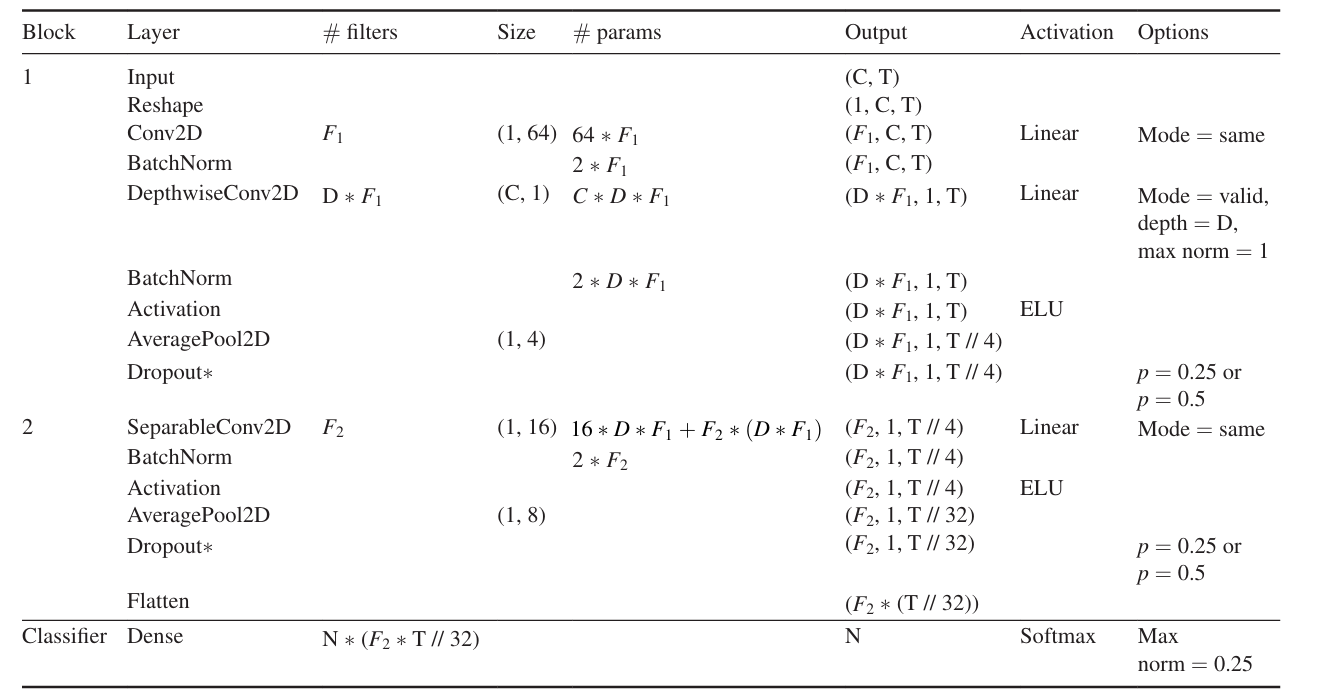} 
    \caption{The architecture of EEGNet \cite{lawhern2018}.}
    \label{table:eegnet_architecture}
\end{table}

\begin{figure}[htbp]
    \centering
    \includegraphics[width=0.8\textwidth]{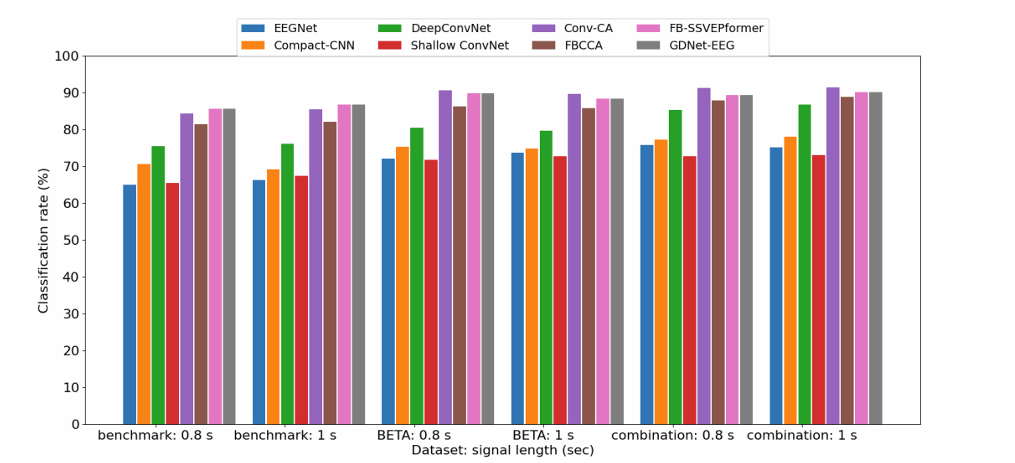} 
    \caption{The average classification accuracies obtained by GDNet-EEG and five other baseline models including EEGNet \cite{wan2023}.}
    \label{fig:eegnet_acc}
\end{figure}

In SSVEP classification tasks, EEGNet is widely used as a baseline model to evaluate the performance of newly proposed models.\cite{wan2023,ding2024} For example,  \cite{wan2023} uses EEGNet as one of the baseline models to demonstrate the efficiency of their GDNet-EEG in SSVEP classification. The model is validated on two public SSVEP datasets (Benchmark and BETA), as well as their combined dataset. With 1-second input signals, Figure~\ref{fig:eegnet_acc} shows the classification accuracies over 10-fold cross-validation using a signal length of 0.8 and 1 s. Similarly to these articles, we will be using classification accuracy on the validation set as a way to evaluate the efficacy of our model extensions. 

Although EEGNet is not the most advanced deep neural network for EEG BCI applications, it is one of the most used and reliable versions. The state-of-the-art models today such as FB-SSVEPformer and GDNet-EEG are all based off of EEGNet, and we used EEGNet as to not confound our proposed model extensions with these other models as they may not be compatible.

In this project, we used EEGNet as a baseline model to evaluate the performance of our proposed methods.

\section{Data}

\subsection{Benchmark Dataset}

For evaluating the performance of our models, there are several well-established and commonly used public SSVEP datasets. The Tsinghua SSVEP BCI speller dataset, also known as the Benchmark dataset, includes EEG recordings from 35 participants using a 40-character speller with flickering stimuli ranging from 8 to 15.8 Hz \cite{wang2017}. There were 64 electrodes used. In the experimental paradigm, 6 blocks of 6 seconds each were tested for each of the 40 targets. The first 0.5 seconds had no stimuli and the later 5.5 seconds did. It is widely used to evaluate SSVEP-based BCI systems due to its controlled setup. 

\subsection{Data Preprocessing}

For data preprocessing, EEG signals are filtered using a Chebyshev Type I filter with cutoff frequencies set between 6 and 90 Hz and stopband corners from 4 to 100 Hz as suggested by  \cite{wan2023}. This is done to eliminate biological artifacts like heart beats and eye blinks which have relatively low frequencies and muscle contractions which can have higher frequencies than neural oscillations. Each trial of filtered multi-channel EEG data is first downsampled from 250Hz to 67.5Hz, and then segmented into smaller chunks, with each segment 4-second length (250 time points), specifically the first 4 seconds of the 6-second trial. The segmentation is also labeled according to the target frequency presented during that trial. Additionally, normalization is applied channel-wise to standardize the EEG data. 

It should be noted that our baseline model cannot be directly compared to that of the previous papers stated as we are using a much longer segment length of 4 seconds to \cite{wan2023}'s 1 second. Note that we maintained the same number of time points as to not increase the complexity of EEGNet. Everything besides the time length is kept the same as \cite{wan2023} to maintain a comparable baseline. Because of the difference in duration of segments, we observed a higher baseline performance as will be shown later, which is expected given the trend of increasing time length resulting in higher accuracy.

The reason we use the first 4 seconds as it is necessary for evaluating our hybrid model as we require the initial change in SSVEP data found at the 0.5 s mark of each trial, which is the trigger for beginning a new character. The SSVEP speller screen will activate on and off as the user spells out a sequence of letters, so the initial trigger is a core part of the SSVEP signal's features. So while \cite{wan2023} is able to cut each trial into 6 1-second segments for training their models, we cannot assume all segments in a trial are identical as they do not include that initial activation that indicates a new character. If we were to only use the first 1 second of data and neglect the rest, then our dataset would be 6 times smaller than the ones used in \cite{wan2023} and would train significantly worse. Therefore, we believed that it would make the most sense to use longer segments with lower sampling rates so that our datasets contain about the same amount of information as the ones in \cite{wan2023}.

\subsection{Data Partitioning}

In the Benchmark dataset, there are data from 35 subjects. After preprocessing, data from each subject has been transformed into small segments. We divided these preprocessed data into a training set, a validation set, and two test sets. In EEG classification, the inter-subject variability is significant, and we aim to separately evaluate performance on subjects whose data are partially used for training and unseen subjects whose data are not involved in training. To achieve this, we selected two subjects (specifically subjects 1 and 9) as the "invisible" test set, with their data completely excluded from training and validation. The remaining 33 subjects' data were randomly split into training, validation, and visible test sets in a ratio of 8:1:1.

\section{Baseline Implementation}

As mentioned in previous sections, we used EEGNet as our baseline model for the SSVEP classification task with the Benchmark dataset (See Figure~\ref{fig:eegnet_acc}). 

The EEGNet model is imported from \href{https://github.com/vlawhern/arl-eegmodels}{GitHub} \cite{lawhern2018}. It is configured with the following parameters: nbclasses=40 specifies 40 output classes, corresponding to the target SSVEP frequencies. The input EEG data has Chans=64 for the 64-channel setup and Samples=250 to represent each trial's 250 time points. A dropout rate of dropoutRate=0.2 is applied to prevent overfitting. Temporal convolution uses a kernel length of kernLength=64 to capture time-dependent patterns, with F1=16 filters for the first layer. Depthwise convolution is configured with a multiplier D=2 to increase spatial feature extraction, followed by F2=32 filters in the second convolutional layer to refine spatial representations further. The model uses standard dropoutType='Dropout' for regularization. 

The model is trained using a batch size of 64 and optimized with the Adam optimizer, initialized at a learning rate of 0.001. To dynamically adjust the learning rate during training, we apply the ReduceLROnPlateau scheduler, configured to monitor validation loss with a reduction factor of 0.5, a patience of 3 epochs, a minimum learning rate of 1e-6. The model is trained for 20 epochs and 100 epochs to evaluate its performance.

The evaluation metric used for the baseline and data augmentations is classification accuracy. The keys that correspond to SSVEP inputs are one-hot encoded. Classification accuracy is calculated by first counting the number of instances where the highest probability output by the model corresponds to the correct one-hot encoded target. This number is then divided by the total number of instances in the dataset to get the classification accuracy. This follows this simple equation: $Acc = \frac{\text{\# of correct predictions}}{\text{total \# of predictions}}$. This is a logical metric to use as this directly corresponds to how accurate the EEGNet model is at distinguishing SSVEP signals. A slightly different evaluation metric for the hybrid model is used and discussed later when the custom dataset is defined.

The loss function used for the baseline, data-augmented models, and hybrid models, is cross-entropy loss. This is because we have a known 40-class target vector and our models produce 40-class probability distributions. The equation for cross-entropy loss is well known and is in Equation \ref{eq:cross_entropy}, where N is the number of samples in a batch, C is the number of classes, y is the known truth and $\hat{y}$ is the predicted label \cite{Shannon1948}. 

\begin{equation}
\mathcal{L} = -\frac{1}{N} \sum_{i=1}^{N} \sum_{c=1}^{C} y_{i,c} \log \hat{y}_{i,c}
\label{eq:cross_entropy}
\end{equation}

In some experiments, we used an alternative loss called a hinge loss which is designed to create a maximum margin between correct and incorrect probabilities. Equation \ref{eq:hinge} shows the hinge loss used as defined by Weston and Watkins for multi-class classification problems, where $\mathbf {x}$ is the input to the model, $\mathbf {y}$ is the output of the model, $\mathbf {t}$ is the target output, and $\mathbf {w}_y$ and $\mathbf {w}_t$ are the representations of the model.

\begin{equation}
    \mathcal{L} =\sum _{y\neq t}\max(0,1+\mathbf {w} _{y}\mathbf {x} -\mathbf {w} _{t}\mathbf {x} )
\label{eq:hinge}
\end{equation}

This baseline model performs very well on the dataset described above. After 20 epochs it has a validation accuracy of 89.19\% and after 100 epochs it has a validation accuracy of 92.97\%. Note that the EEGNet in \cite{wan2023} achieved about 65\% validation accuracy on 1-second trials, which means our model did expectedly better as we are using 4-second trials. The explanation for this discrepancy was explained in the data preprocessing section. Overall, this result was expected and proves our baseline model is training correctly. This is the baseline performance that we attempt to improve in this paper. 

\section{Model Extensions}

\subsection{Data Augmentation}

One ongoing struggle in the field of BCI is generalizing learned models to individuals not in the training data. Everyone's brain is wired differently and has various electrical characteristics, all of which make it hard for a model to generalize well. One area of research which is promising on this front is that of creative data augmentation that uses domain knowledge and expertise to help in generalization. There are many possible augmentation techniques for SSVEP data, but we will list a few here. In \cite{ding2024}, the authors utilize mask encoding to force the model to learn more robust features. In this paper, the authors only mask in the time domain, but it could also be worth looking into masking in the frequency domain. In \cite{mai2023} a different approach is taken in which the authors build off of previous work in which time-shifted sections are used in training (essentially time-cropping). In this paper, the amount of shifting was determined intelligently for each SSVEP class based on the local similarities of data in a single recording.

In another example, \cite{pan2024} augments SSVEP data using two methods, both of which are done in the frequency domain after applying an Fast Fourier Transform (FFT). The first is filter band masking which is essentially applying a notch filter to the data. The second is random phase erasing in which random sections of the frequency data are phase-shifted to a random phase, then the data is returned to the time domain. This is a very interesting concept to us as it attempts to recognize that the time that a signal takes to reach an electrode varies significantly on the person. One weakness we see in this approach is the fact that, in reality, the phase shift of frequency components is not random, but determined by the physical characteristics of the brain and skull.

Our target in this section of our study was to investigate various data augmentation techniques and analyze how the models generated using them perform to the baseline model. Below are the data augmentations we applied to the SSVEP Benchmark dataset and the reasoning behind them.

\begin{itemize}
    \item \textbf{Frequency Masking:} The idea behind frequency masking is that there is useful information found in every part of the frequency spectrum. However, some portions of the spectrum may be more impactful. Thus, they are weighted higher during regular gradient descent. This results in the weights that deal with the impactful sections being learned quickly while the weights dealing with the less impactful sections wander and do not converge. By masking random sections of the frequency domain during training, the model is forced to learn other weights when it does not have access to very impactful portions of the data.
    \item \textbf{Time Masking:} The concept behind this data augmentation technique is identical to the previous technique except the masking is done in the time domain rather than in the frequency domain.
    \item \textbf{Random Phase Noise:} This is the augmentation described above in which random portions of the frequency domain are given a random phase adjustment. The idea behind this augmentation is that different people's brains behave differently. Thus, it is reasonable to think that various frequency signals will come at slightly different times, equating to phase adjustments in the signals.
    \item \textbf{Random Magnitude Noise:} In this augmentation, the same thing is done as random phase noise, except the noise is added to the magnitude of the data instead of the phase. The reasoning behind this is similar, attempting to generalize the data in ways that would be expected in other people's brains.
    \item  \textbf{Random Impulse Addition:} The idea behind this augmentation is that brain signals are electrical with complex signal dynamics. This means there are reflections that could show up as "echos" in the EEG data. So, for this augmentation, we add random echos to the EEG data. This is an attempt to create a more generalizable model.
    \item \textbf{Random Salt and Pepper Noise:} The idea behind this augmentation is to make the model robust to noise in the sensing equipment. We add random values to random points in the time domain to simulate imperfect signal measurement. 
\end{itemize}

Each of these augmentation techniques were tested via an ablation study on the Benchmark dataset.

\subsection{Language Model}

One area of novel experimentation our group pursued in this project is improving the classification of SSVEPs for the BCI speller task by incorporating letter frequency into the character classification process. Currently, SSVEP-based spellers only use the spectral, temporal, and spatial features of the EEG data to determine a character's identity, but while that would be the only option if the user is looking at arbitrary stimuli, we know that the speller task contains other useful information. If we assume the user is trying to spell English words, we know that if they first spell the letter 'Q', we can be quite confident that the letter 'U' will immediately follow. Therefore, we can assign a higher weight to the probability of the letter 'U' after confirming a 'Q'. This should improve the accuracy of the SSVEP speller especially when a specific trial has low confidence in EEGNet.

\begin{figure}[htbp]
    \centering
    \includegraphics[width=0.8\textwidth]{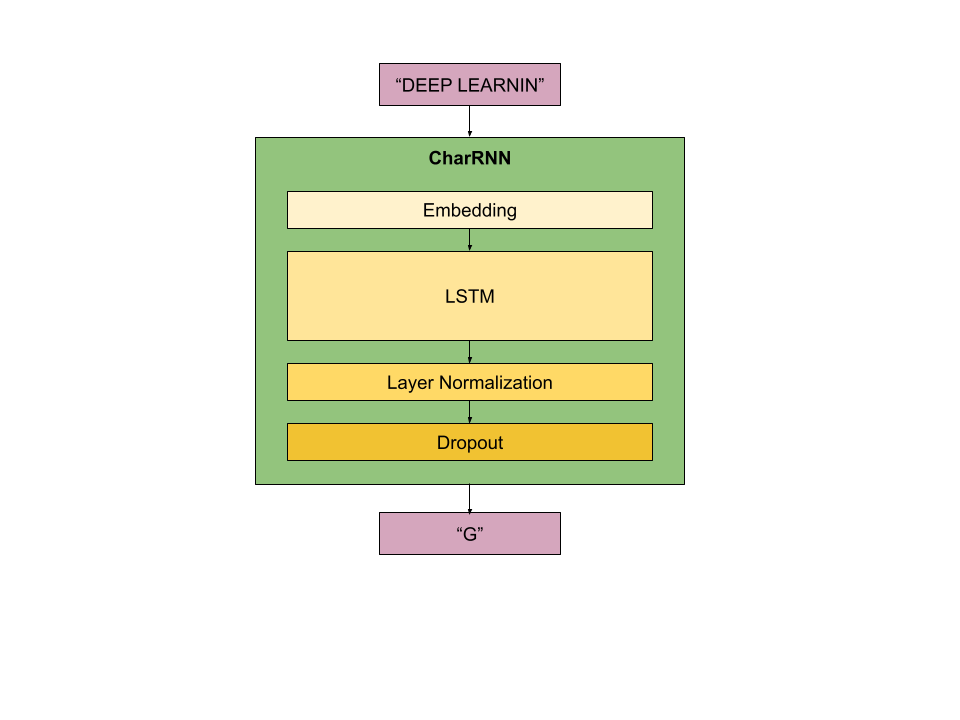} 
    \caption{CharRNN Diagram}
    \label{fig:charRNN_diagram}
\end{figure}

In order to incorporate this letter frequency information into EEGNet, we implemented a character-based RNN as developed by \cite{graves2013}. This RNN would work by sequentially processing each character in a text, using each character as input while updating its hidden state to capture context from previous characters, allowing the model to predict future letters. The architecture can be seen in Figure \ref{fig:charRNN_diagram}. The inclusion of layer normalization and dropout helps CharRNN to have more stable learning and generalization. When given a string as input, CharRNN provides a prediction of the subsequent letter.  

A more detailed summary of CharRNN is in Table \ref{tab:char_rnn_summary}. CharRNN is trained on WikiText2, a large text corpus from hand-picked, high-quality Wikipedia articles, which provides a good source of accurate letter frequency information. The batch size is 128, the learning rate is 0.001, the loss is Cross Entropy Loss, and we used the Adam optimizer and a CosineAnnealing scheduler. Only 3 epochs were necessary to reach a stable plateau. Note that the vocabulary size is 27 which are the 26 capital letters from A to Z and also 'space'. Our EEGNet model is trained on a 40-class dataset so we will be cutting the output of EEGNet to the 26 classes that correspond to the letters of the alphabet (which in this case are targets 0-25).

\begin{table}[h!]
\centering
\begin{tabular}{|l|c|c|}
\hline
\textbf{Layer (type:depth-idx)} & \textbf{Output Shape} & \textbf{Param \#} \\ \hline
\textbf{CharRNN}                & {[}128, 20, 27{]}    & --               \\ \hline
Embedding: 1-1                  & {[}128, 20, 128{]}   & 3,456            \\ \hline
2-Layer LSTM: 1-2                       & {[}128, 20, 256{]}   & 921,600          \\ \hline
LayerNorm: 1-3                  & {[}128, 20, 256{]}   & 512              \\ \hline
Dropout: 1-4                    & {[}128, 20, 256{]}   & --               \\ \hline
Linear: 1-5                     & {[}128, 20, 27{]}    & 6,939            \\ \hline
\end{tabular}
\caption{CharRNN Model Summary}
\label{tab:char_rnn_summary}
\end{table}

\begin{figure}[htbp]
    \centering
    \includegraphics[width=0.8\textwidth]{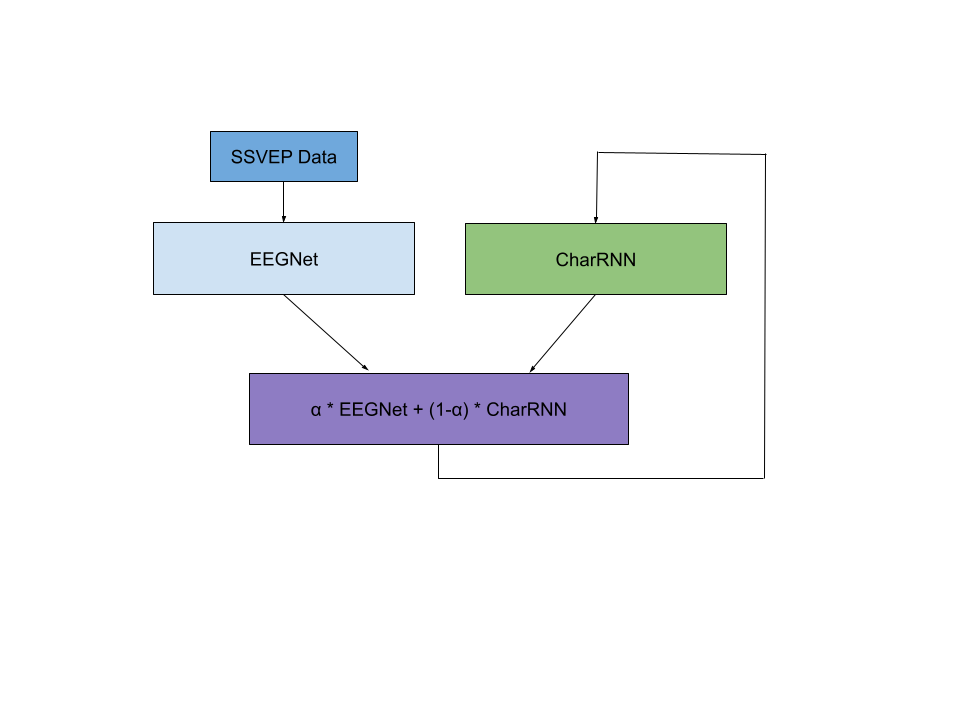} 
    \caption{Hybrid EEGNet-CharRNN Model}
    \label{fig:Hybridmodel_diagram}
\end{figure}

The CharRNN is combined with EEGNet to create our hybrid model EEGNet-CharRNN, as seen in Figure \ref{fig:Hybridmodel_diagram}, where the SSVEP data is taken in by EEGNet and previously-predicted characters are taken in by CharRNN, and the output of the model (which is the next predicted character), is determined using Equation \ref{eq1}. Here, EEGNet(C) is the probability distribution from EEGNet and CharRNN(C) is the probability distribution of CharRNN. Essentially, when the subject looks at their first character in the speller screen, the SSVEP data is taken and processed by EEGNet alone. This predicted letter is sent to CharRNN, which prepares its prediction for the second letter. Then when the subject looks at their second character, that SSVEP data is processed and the predictions of both models are combined when choosing the next character, which will then go back into CharRNN for future character predictions. This loop repeats until the word terminates.

In the evaluation of this model, various $\alpha$'s are tested to determine the best combination of the two models.

\begin{equation}
P(C) = \alpha * EEGNet(C) +  (1-\alpha) * CharRNN(C)
\label{eq1}
\end{equation}

\subsubsection{Dataset Creation}

The Benchmark dataset's original experimental paradigm only consisted of testing individual targets as opposed to entire words/sentences. Therefore, in order to properly evaluate our hybrid model, we created our own words by concatenating random trials of corresponding characters. The choice of specific trials for each character in a word is random to mimic the expected SSVEP signal from a participant spelling a word in a real experiment setting. This was done with all subjects in the Benchmark dataset, using the same preprocessing steps used for EEGNet training. Importantly, we are keeping the first 4 seconds of each trial as this contains a 0.5 second "rest" period needed to express separations between letters. In total, three datasets were created:

\begin{enumerate}
\item Top 100 Common: The top 100 most common English words according to Google Web Trillion Text Corpus. This is a logical choice as a subject would be expected to spell a lot of these important words.
\item Top 100 Long: The top 100 most common English words that are at least 7 letters long, according to Google Web Trillion Text Corpus. The idea is that CharRNN would perform better the longer the words are as the prediction confidence would increase.
\item Top 1000 Common: The top 1000 most common English words according to Google Web Trillion Text Corpus. This is just a larger version of the first dataset to see if the ability of the hybrid model extends to more words.
\end{enumerate}

\subsection{Hybrid Model Evaluation Metric}

To evaluate the hybrid model on these datasets, we will use the classification accuracy of each individual character out of all characters in the dataset, which follows this simple equation: $Acc = \frac{\text{\# of correct characters}}{\text{total \# of characters}}$. This makes the most sense as an increase in per-character accuracy leads to a greater level of meaningful information being transferred through the speller.

\section{Results and Discussion}

\subsection{Data Augmentation}

\begin{figure}[ht!]
    \centering
    \includegraphics[width=\linewidth]{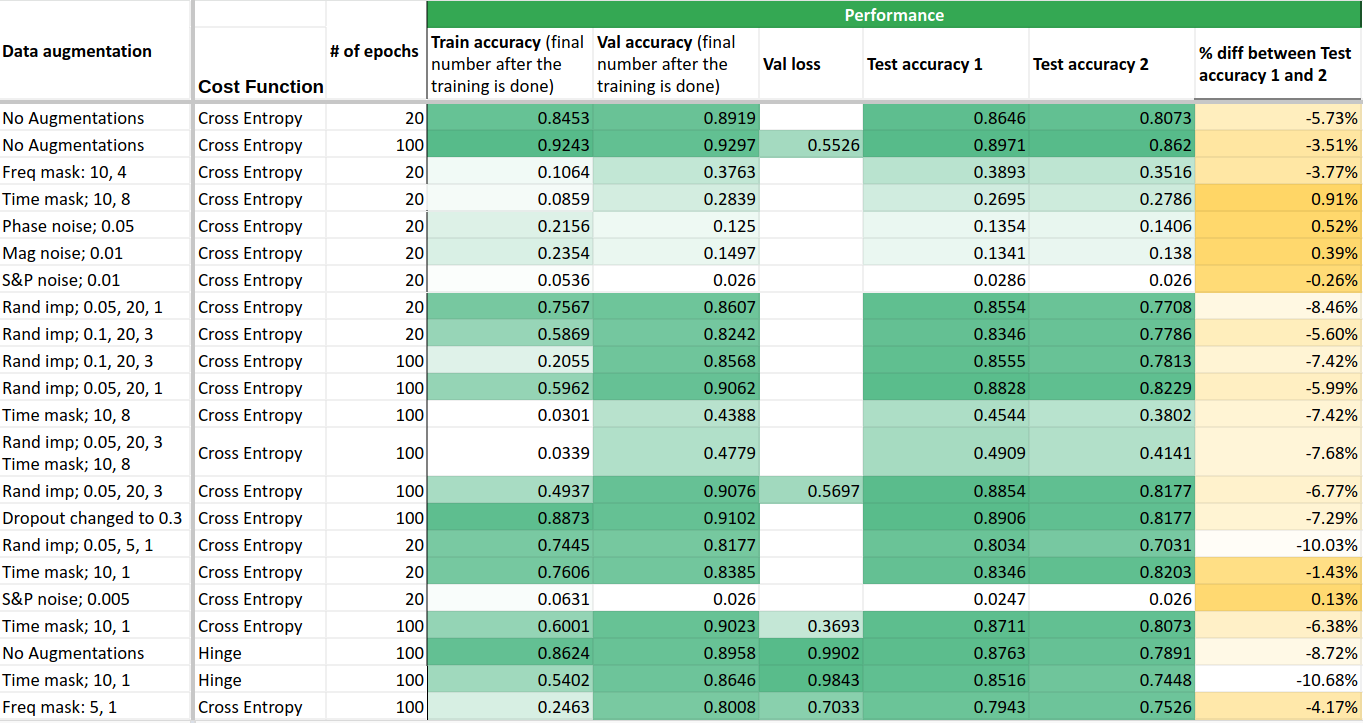}
    \caption{Table of ablations}
    \label{fig:abltations-table}
\end{figure}

The results of using the various data augmentation techniques during training of EEGNet on the Benchmark dataset were generally disappointing, but there are several interesting things to note. The ablation table is shown in Figure \ref{fig:abltations-table}. In the figure, "Test accuracy 1" refers to the classification accuracy of a given model on the "visible" test set while "Test accuracy 2" refers to the classification accuracy of a model on the "hidden" test set.

For the first ablations, we did not record the validation loss, only the accuracies of the model on the various datasets. After a bit, we began recording the validation loss because we found it held valuable insight into the effect of augmentations. Overall, the best validation and test accuracy was trained by using no augmentations and running for 100 epochs. All of the data augmentation techniques resulted in decreased performance, some more than others. Salt and pepper noise was by far the worst, ultimately causing the model to learn nothing of substance. We hypothesize this is due to the fact that adding noise to every timestep, no matter how little makes it much harder for a model to extract the underlying information. Phase noise and magnitude noise performed poorly as well. Interestingly, these two are the only two in which the validation and test accuracies are lower than the training accuracy. Usually, it is the other way around due to dropout. We believe this phenomenon is caused by the model learning to identify data in the test dataset that has been altered, but this alteration or augmentation does not correspond to the generalized differences between the training and validation/test datasets.

Random impulse addition showed early promise because it did not severely negatively impact the validation accuracy. There was an early hypothesis that the model just needed to train longer. However, this hypothesis was disproved after several further ablations with more epochs. A key thing to notice about this augmentation is that it negatively impacted the generalizability of the model. The difference between the first and second test accuracies increased in a small but statistically significant way. We draw a similar conclusion as we did with random phase and magnitude noise. The augmentation is simply not a good approximation of how EEG changes over time and between subjects.

Frequency and time masking were the most promising data augmentation techniques. While frequency masking never improved upon the baseline in either accuracy or loss, the training accuracy was very low while maintaining a very high validation loss. This implies that there is room to alter the hyper-parameters to improve the baseline model using frequency masking. Improving the baseline model was achieved using time masking. While the validation and test accuracies did not improve using time masking, the validation loss did improve significantly. This improvement in the loss could have a valuable impact if EEGNet were used as part of a larger system such as the hybrid model we discuss next.

One other interesting and important thing to note is the difference between the visible test set accuracy and the hidden test set accuracy in each of the ablations. The baseline model has a difference of 3.5\% between the two, suggesting the model generalizes well to new subjects. Some augmentations, such as time masking after 20 epochs result, in a reduced difference between the two (1.4\%). While this is not statistically significant, there is promise there for time-masking to improve generalizability. In every case with a high accuracy, the model performs better on the visible dataset than it does on the hidden dataset. This makes sense because it is logical for it to perform better on data more similar to what it was trained on.

To understand the table, here is the key for the parameters used in each augmentation.

\begin{itemize}
    \item \textbf{Freq mask; x, y:} where x is the maximum width of the masks and y is the maximum number of masks applied to each input.
    \item \textbf{Time mask; x, y:} where x is the maximum width of the masks and y is the maximum number of masks applied to each input.
    \item \textbf{Phase noise; x:} where x is the standard deviation of the noise applied to the phase component of the input.
    \item \textbf{Mag noise; x:} where x is the standard deviation of the noise applied to the magnitude component of the input.
    \item \textbf{S\&P noise; x:} where x is the standard deviation of the noise applied to the input at each time step.
    \item \textbf{Rand imp; x, y, z:} where x is the maximum scaling of each echo, y is the maximum time delay of each echo, z is the maximum number of echos added to each input.
\end{itemize}

\subsection{Language Model}

\begin{figure}[htbp]
    \centering
    \includegraphics[width=0.8\textwidth]{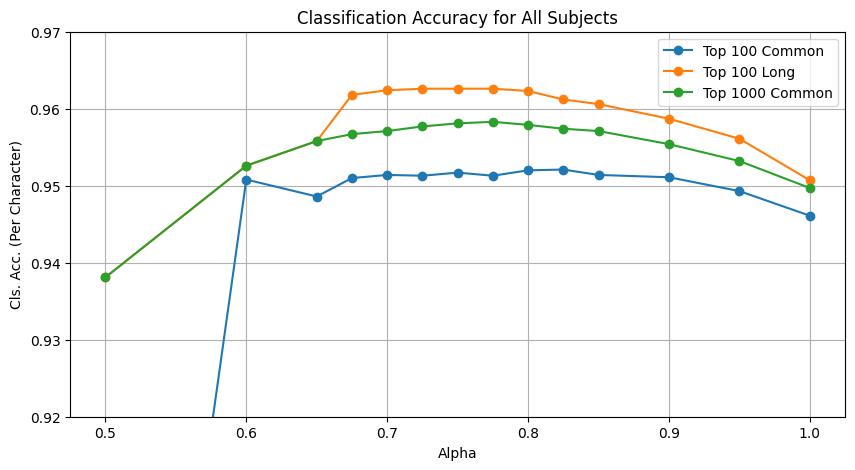} 
    \caption{Hybrid Model Evaluation on All Subjects}
    \label{fig:hybridresults_all}
\end{figure}

After running several ablations of $\alpha$ on the hybrid model for the three datasets, we get the results in Figure \ref{fig:hybridresults_all}, which is accuracy when tested on all subjects.

The baseline model performance of just EEGNet is seen at the right when $\alpha = 1.0$, and it already performs quite well on all three datasets with about 95\% accuracy. This is expected since we are using all subjects, and our baseline model has a high training accuracy so we expectedly can get most SSVEP data correct.

As the $\alpha$ is decreased, meaning more of the CharRNN is used for predicting characters, the accuracies go up for all 3 datasets, peaking at around 96\% accuracy at an $\alpha$ of 0.75. This is an increase of 0.6\% to 1.19\% depending on the dataset. Although a small increase, this is quite significant considering the high baseline, and it demonstrates the hybrid model's potential at improving SSVEP speller efficiency. The reason this works is simple: when EEGNet is not confident about a prediction, CharRNN can supplement it with a high-confidence prediction based on previously predicted letters. This is especially effective when EEGNet already predicts several correct letters as that boosts the confidence of CharRNN significantly. This exact inference can be seen when observing letters that CharRNN fixes, which are usually when confidence from EEGNet is low.

Furthermore, we can observe that the accuracy plummets when $\alpha$ gets closer to 0.5, which makes sense as by then, we would be primarily relying on CharRNN for predictions instead of EEGNet, and without a good starting substring to base their predictions off of, CharRNN won't provide meaningful outputs. While this may work for some of the more common words, less common ones will not be decoded correctly.

\begin{figure}[htbp]
    \centering
    \includegraphics[width=0.8\textwidth]{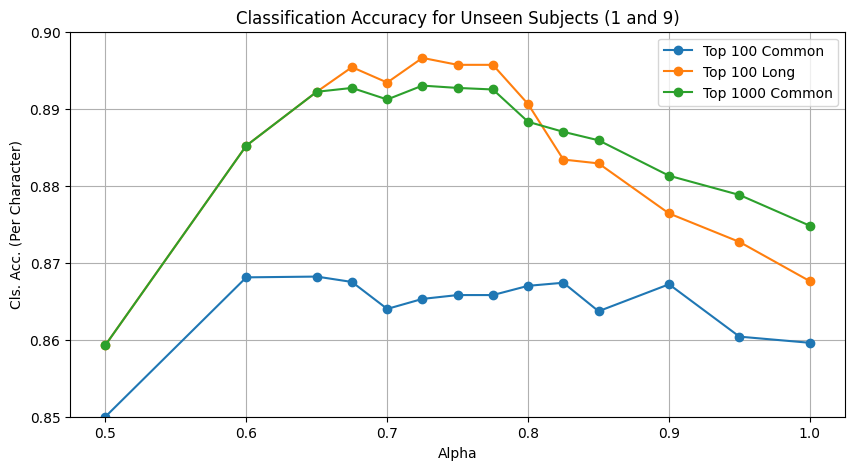} 
    \caption{Hybrid Model Evaluation on Unseen Subjects}
    \label{fig:hybridresults_unseen}
\end{figure}

In Figure \ref{fig:hybridresults_unseen}, we see the classification accuracies for the two unseen subjects 1 and 9, which were hidden from the training dataset for EEGNet. The baseline accuracy is much lower as expected since EEGNet struggles with generalizing to new data, with values at about 87\%. This is where the hybrid model shows the most benefit, achieving up to 2.9\% increased accuracy at an $\alpha$ of about 0.75. These results make sense since the more uncertain EEGNet is, the greater the benefit is provided by CharRNN. Considering that generalizability is one of the main problems with EEG classification today, this demonstrates the potential for hybrid models to improve accuracy on specific domains.

Furthermore, there is uncertainty with these results is that we are working with artificially-stitched SSVEP data, and it is possible that real-time SSVEP data corresponding to full words/sentences looks different than our dataset. Also, a more proper evaluation of the hybrid model would be allowing the subjects to write whatever they wanted as this would give us user intent. Due to limited access to SSVEP speller setups, we could not perform these experiments ourselves and had to rely on creating artificial words out of prior SSVEP datasets.

\section{Future Work}

This project serves as a starting ground for experimenting with data augmentation and language modeling to improve SSVEP spellers. There is a lot of additional research to be done, such as exploring different parameters for time masking. However, the largest contribution that comes from the data augmentation work that we have done is a starting place for future researchers to begin from, aware of which data augmentation techniques have potential and which ones they should not waste their time on.

For the hybrid model with language modeling, testing the model in real-time with experimental paradigms that emphasize communicating full sentences would allow for a more practical evaluation of the model that would mimic real-life usage. The way that the hybrid model is currently set up allows for real-time processing without increased latency as long as the modules are run in parallel, so this would be a viable approach to extending the analysis. Furthermore, larger language models such as transformers would be helpful for improving predictions when conducting these full-sentence tests, as they would perform much better than our character-based RNN. Finally, combining the best-performing data augmentations with the best-performing language models, in addition to other methods, would allow us the attain the best model for practical real-time SSVEP speller decoding.

\section{Conclusion}

In this paper, we attempted to improve the performance of a common model for EEG data for the SSVEP speller task. The problem with many EEG deep learning models today is their inability to learn complex features present in the data and leverage them to produce a high-accuracy classification. This problem is exacerbated when the models try to generalize to unseen subjects. In this work, we went about improving the accuracy of a baseline SSVEP speller based on EEGNet in two ways. First, we experimented with various data augmentation techniques with the hopes that one or several of them would improve accuracy on either visible or hidden subjects. Second, we coupled our best EEGNet model with an RNN language model to predict the next characters in a sequence using inputs from both models.

The data augmentations we used proved to be largely unfruitful. The one exception to this is time masking which had a significant improvement over the baseline in the validation loss, although it did not improve the validation or either test accuracies.

The combination of the best EEGNet model with a character RNN resulted in a measurable improvement in accuracy. The addition of the language model especially improved the accuracy for hidden subjects, making a difference of 2.9\%. The improvement in unseen subjects using the hybrid model can be attributed to the ability of the language model to make up for EEGNet's weaknesses. Generally speaking, the poorer the performance of EEGNet, the larger the difference a language model can make. Thus, for unseen subjects, when EEGNet performance is comparatively poor, the character RNN has a greater positive impact.

The findings of this paper are valuable and novel contributions to the field; with these contributions to the field come increased hope for this technology to help those who need it. A comprehensive investigation into various data augmentation techniques gives future researchers a place to stand as they move the field forward. Our novel combination of EEGNet with a language model shows great potential for real technology to develop based on SSVEP spellers and is supported by a large language model for automatically filling in likely thoughts for the user.

\section{Work Distribution}

All team members participated in every aspect of the project, however, we list below the tasks that each team member led and spent significant time working on.

\begin{itemize}
\item Joseph: Implemented the hybrid EEGNet-CharRNN model. Created the word-based dataset for evaluating the hybrid model. Ran ablations on the hybrid model.
\item Ruiming: Implemented data preprocessing and baseline EEG model.
\item David: Implemented data augmentation and ran ablations for it on the Benchmark dataset.
\item Kipngeno: Helped implement baseline EEGNet model.
\item Kateryna: Advised on deep learning model design, evaluation strategy, and project direction.
\end{itemize}

\section{Code}

All of the code for this project can be found on \href{https://github.com/kkipngenokoech/Hybrid-EEGNET-CharRNN-predictor}{our public GitHub page}.

\bibliographystyle{plainnat} 
\bibliography{arxiv_submission}   
\end{document}